%% file: domInter.tex
\documentclass[conference]{IEEEtran}
\IEEEoverridecommandlockouts

\usepackage{cite}
\usepackage{amsmath,amssymb,amsfonts}
\usepackage{dsfont}

\usepackage{algorithmic}
\usepackage{graphicx}
\usepackage{textcomp}
\usepackage{xcolor}
\usepackage{multicol}

\usepackage{geometry}
\newcommand{\norm}[1]{|\!| #1 |\!|}
\newcommand{\Ii}{\mathcal{I}}
\newcommand{\al}{\alpha}
\newcommand{\Ll}{\mathcal{L}}

\newtheorem{thm}{Theorem}

\newtheorem{cor}{Corollary}[thm]

\graphicspath{{./images/}}

\title{On Dominant Interference in Random Networks and Communication Reliability}

\author{
  \IEEEauthorblockN{Dong Liu,
  Baptiste Cavarec,
  Lars~K. Rasmussen,
  and~Jing Yue}
\IEEEauthorblockA{\textit{Division of Information Science and Engineering} \\
  \textit{KTH Royal Institute of Technology}\\
  Stockholm, Sweden \\
  E-mail: \{doli, cavarec, lkra, jyue\}@kth.se}
\thanks{This research was supported by the Swedish Research Council.}
}

\def\BibTeX{{\rm B\kern-.05em{\sc i\kern-.025em b}\kern-.08em
    T\kern-.1667em\lower.7ex\hbox{E}\kern-.125emX}}

\begin{document}

\maketitle

\input{section/sec-abstract}
\input{section/sec-intro}
\input{section/sec-model}
\input{section/sec-application}

\input{section/sec-numerical}
\input{section/sec-conclusion}

\appendices
\input{section/sec-appendix}

\bibliographystyle{IEEEtran}
\bibliography{domRef}

\end{document}

%% file: section/sec-abstract.tex
\begin{abstract}\label{abstract}
   In this paper, we study the characteristics
  of dominant interference power with directional reception in a random
  network modelled by a Poisson Point Process. Additionally, the Laplace functional of cumulative
  interference excluding the $n$ dominant interferers is also derived,
  which turns out to be a generalization of omni-directional reception and complete accumulative interference. As an application of these results, we study the impact of directional receivers in random networks in terms of
  outage probability and error probability with queue length constraint.

\end{abstract}

%% file: section/sec-intro.tex
\section{Introduction}
The studies of communication technologies,
services and applications for massive Machine Type Communication (MTC)
have significantly increased in recent years. Emerging MTC requires significant performance improvements for a communication link, which is either an increase of capacity for vehicular technology, or an increase in high reliability while keeping low latency constraints for tactile internet and factory automation purposes. 
A key point of these challenges is to accommodate the increasing number of devices with reliable services.

Multiple access schemes, especially non-orthogonal multiple
access, have been identified as potential solutions to meet future
massive access requirements in wireless networks under various applications
\cite{shirvanimoghaddam2017massive,ding2015cooperative,ma2015performance,wildemeersch2013successive}.
However, fading is modeled as path loss, and hence
  only the Euclidean distance was taken into account to derive the
  $n^{\text{th}}$ ($n=1,2,\cdots$) dominant interference power in
  these works.
   A closed form of the Euclidean distance
distribution to the $n^{\text{th}}$ nearest
neighbor in networks modeled by a Poisson Point Process (PPP) was derived
in \cite{haenggi2005distances, haenggi2012stochastic}.

Though path loss is one of the major factors
causing wireless signal fading, channel fading is also significantly influencing the strength of wireless signals. 
Due to the dynamic nature of such fading, the signal power
received from a nearer transmitter is possibly smaller than that from
a farther transmitter. Thus we want to study
characteristics of the $n^{\text{th}}$ dominant
interference power as well as accumulative
interference in terms of power itself rather than Euclidian distance. It is also of
interests to see how these affect communication reliability in
wireless networks.
In addition, directional reception induces sectorization in cellular networks, which is a typical method for interference management\cite{psomas2017impact,yi2003capacity}. Thus, directional reception angle also affects the characteristics of dominant and accumulative
interference power.


In this paper, we model stationary random networks by
homogeneous PPP. We give probability distribution of the $n^{\text{th}}$ dominant
interference power in stationary random networks where receivers have directional
reception. In addition, the
partial accumulative interference excluding some dominant interferers is studied,
which is a generalization of the omni-directional case and also complete
accumulative interference. The outage probability and transmission
error probability with queue length constraint in Nakagami-m fading
are studied.

The remainder of this paper is organized as follows. In
Section~\ref{sec:mainResult}, we derive the closed-form distribution
of the $n^{\text{th}}$ dominant interference power and Laplace functional of
partial accumulative interference power. In Section~\ref{sec:app}, we discuss the
communication outage probability and transmission error probability
with queue length constraint. In Section~\ref{sec:numResult}, numerical results are given to verify the derived results. Finally, we conclude this paper in Section~\ref{sec:conclusion}.


%% file: section/sec-model.tex
\section{Dominant and accumulative Interference}\label{sec:mainResult}

We model a random network by a PPP $\Phi=\{x_i \}$, $x_i
\in \mathbb{R}^2$, with intensity $\lambda$, where $x_i$ denotes the 
coordinates of node\footnote{Term ``node'' and ``point'' are alternatively used depending
  on the context of network or point process.} $i$. The intensity
measure of $\Phi$ is denoted by
$\Lambda$. For a receiver located at $x\in \mathbb{R}^2$ in the
random network, the interference power $I_i$
from node $i$ is

\begin{equation}
  I_i = h_i \norm{x-x_i}^{-\alpha},
\end{equation}
where $h_i$ is the channel gain from $x_i$ to $x$,
$\norm{\cdot}$ the Euclidean norm, and $\alpha >2$ the path-loss
parameter. We arrange $\{I_i\}$ into a
descending-ordered sequence $\left( I_n \right)_{n=1,2, \cdots}$,
such that $I_n$ is the \textit{$n^{\text{th}}$ dominant interference power}
($I_1 >I_2 >  \cdots$).

To study the distribution of $I_n$, point
process mapping and displacement
\cite{haenggi2012stochastic} are be applied on $\Phi$. We assume directional reception and the
reception is within deterministic angle $\phi$, where $0< \phi \leq 2
\pi$. Since $\Phi$ is homogeneous, without losing generality, we define the
interferers to node $x$ by
\begin{equation}\label{def_phi_prime}
  \Phi^{\prime}= \left\{x_i^{\prime}=x_i \mid \angle{\overrightarrow{
        x x_i}} \leq \phi, x_i\in \Phi \right\},
\end{equation}
where $\angle{\overrightarrow{x x_i}}$ gives the angle of vector
$\overrightarrow{x x_i}$ regarding the horizontal axis on $\mathbb{R}^2$.
Then we map $\Phi^{\prime}$ to $\Phi_1$ as
\begin{equation}\label{def_phi1}
  \Phi_1 = \left\{y_i= \norm{x_i^{\prime}-x}^{\alpha} \mid x_i^{\prime} \in
    \Phi^{\prime}, \alpha>2 \right\}.
\end{equation}

We further define the point process $\Phi_2$ that takes values from $\mathbb{R}^{+}$ as a displacement of $\Phi_1$ as
\begin{equation}
  \Phi_2 =   \left\{ z_i=\frac{y_i}{h_i} \mid y_i \in \Phi_1\right\}.
\end{equation}
We arrange
the points of $\Phi_2$ into an ascending-ordered sequence $\left(  z_n\right)_{
  n=1,2, \cdots}$ ($z_1<z_2< \cdots$), along the power axis $\mathbb{R}^{+}$. 
Hence, $z_n$ is the $n^{\text{th}}$ closest point of $\Phi_2$ to origin on the power
axis $\mathbb{R}^{+}$ leading the $n^{\text{th}}$ dominant interference
power to be $ I_n=z_n^{-1}$. 

\begin{thm}[The $n^{\text{th}}$ Dominant Interference Power]\label{theorem1}
  In a PPP on $\mathbb{R}^2$ with intensity
  $\lambda$, $z_n$, the inverse of the $n^{\text{th}}$ dominant interference power
  $I_n$, to a receiver
  with reception angle $\phi$ has the
  Probability Density Function (PDF)
  \begin{equation}\label{pdfzn}
    f^{\phi}_{z_n}(z) = \frac{2 \exp\left\{-\frac{1}{2} \bar{h}_{{2}/{\alpha}} \phi  \lambda z^{2/\alpha }\right\} \left(\frac{1}{2} \bar{h}_{{2}/{\alpha}} \phi  \lambda z^{2/\alpha }\right)^n}{\alpha  z (n-1)!},
  \end{equation}
  where $\bar{h}_{2/\alpha} = E_h[h^{2/\alpha}]$ denotes expectation supported by PDF of channel gain $h$.
\end{thm}
\begin{IEEEproof}
  Combining the definition in \eqref{def_phi_prime} and
  \eqref{def_phi1} gives
  \begin{equation}
    \Phi_1 = \{y_i=
    ||x_i-x||^{\alpha} \mid \angle{\overrightarrow{ x x_i}} \leq \phi, \alpha>2,
    x_i \in \Phi \}.
  \end{equation}
  Denote the intensity measure of $\Phi_i$ by $\Lambda_i$ and its
  intensity function by $\lambda_i$. Since $\Phi_1$ is obtained by thinning and mapping
  independently from $\Phi$, $\Phi_1$ is also a PPP
  \cite{haenggi2012stochastic}, \cite{NET-006}. The intensity
  $\Lambda_1$ is 
  \begin{align}
  \!\!\!\!\! \Lambda_1(y)\!=\!E[\Phi_1([0,y])] &=\!\!\!\!\int_{B_x(y)}
                                    \!\!\!\!\!\!\!\!\!\!\lambda
                                    \mathds{1}\left(\angle{\overrightarrow{x x^{\prime}}} \leq \phi\right) dx^{\prime}
                                                    = \frac{\phi \lambda y^{2/\alpha}}{2},
  \end{align}
  where $B_x(r)=\{x^{\prime} \mid ||x-x^{\prime}||<r, x, x^{\prime} \in \mathds{R}^2,r\in
  \mathds{R}^{+}\}$ is a disc centered
  at $x$ with radius $r$, and $\mathds{1(\cdot)}$ is the indicator function.
  Thus, the intensity function of $\Phi_1$ is
  \begin{equation}
    \lambda_1(y) = \frac{\partial{\Lambda_1(y)}}{\partial{y}} =
    \frac{\phi \lambda y^{2/\alpha -1}}{\alpha}, y>0.
  \end{equation}

  Since the point process $\Phi_2$ is actually a displacement of $\Phi_1$,
  $\Phi_2$ is also PPP according to the Displacement Theorem
  \cite{NET-006}. The intensity measure $\Lambda_2$ of $\Phi_2$ can be
  obtained according to the Displacement Theorem. We have
  \begin{equation}
    \mathbb{P}\left(\frac{y}{h}<z\right)=1-F_h\left(\frac{y}{z}\right),
  \end{equation}
  where $F_h(\cdot)$ is the Cumulative Distribution Function (CDF) of
  $h$. The probability kernel $\rho(y,z)$ of the displacement
  is
  \begin{equation}\label{kernalFun}
    \rho(y,z)=\frac{\partial}{\partial{z}}\left[ 1-F_h\left(\frac{y}{z}\right)\right] =
    \frac{y}{z^2}f_h\left(\frac{y}{z}\right).
  \end{equation}

  Hence, the intensity function $\lambda_2(z)$ of $\Phi_2$ is obtained as
  \begin{align}\label{intensity2}
    \lambda_2(z) &= \int_0^{\infty} \lambda_1(y) \rho(y,z) dy = \frac{\phi \lambda}{\alpha} z^{2/\alpha-1}E_h[h^{2/\alpha}],
  \end{align}
  where $E_h[h^{\frac{2}{\alpha}}]= \int^{\infty}_{0}h^{\frac{2}{\alpha}} f_h(h) dh$. Then
  \begin{equation}\label{intensity2final}
    \Lambda_2(z) = \int_0^{z} \lambda_2 (z) dz = \frac{\phi
      \lambda}{2} E_h[h^{2/\alpha}] z^{2/\alpha}, z \in
    \mathbb{R}^+.
  \end{equation}
  
  We arrange
  points of $\Phi_2=\left\{z_n \right\}$ into an ascending-ordered
  sequence $\left( z_n \right)_{n=1,2,\cdots}$. By definition,
  $I_n=z_n^{-1}$ is the $n^{\text{th}}$ largest interference power from points of $\Phi_2$. Denote the number of points of
  $\Phi_2$ within $\left[ 0, z\right]$ on the power axis
  $\mathbb{R}^{+}$ by ${N}_z$. Then ${N}_z$ is a
  Poisson random variable with intensity $\Lambda_2(z)$
  and we have
  \begin{equation}
   \!\! \mathds{P}({N}_z=k) =
    \frac{\left(\Lambda_2(z)\right)^k}{k!} \exp\{-\Lambda_2(z)\}, k=0,1,2,\cdots.
  \end{equation}
  Let $F_{z_{n}}^{\phi}(z)$ be the CDF of $z_n$, we have
  \begin{align}
    F_{z_n}^{\phi}(z)  = 1-\sum_{k=0}^{n-1}\mathds{P}({N}_z=k) = \frac{\gamma\left(  n, \Lambda_2(z) \right)}{\Gamma(n)},
  \end{align}
  where $\Gamma(a)=\int_0^{\infty}x^{a-1} \exp\{-x\} dx$ is gamma function
  and $\gamma(a,b)=\int_0^{b}x^{a-1} \exp\{-x\} dx$ is lower incomplete
  gamma function.  By taking the derivative of $F_{z_n}^{\phi}$ the PDF of $z_n$ is
  \begin{equation}
    f_{z_n}^{\phi}(z)=\frac{2 \exp\{-\Lambda_2(z)\} \left( \Lambda_2(z)
      \right)^n}{\alpha \Gamma(n) z}.
  \end{equation}
  
\end{IEEEproof}

\begin{cor}
  By Theorem~\ref{theorem1}, the $n^{\text{th}}$ dominant interference power is expected to be
  \begin{equation}
    E[I_n]= \left( \frac{\phi  \bar{h}_{2/\alpha} \lambda}{2}
    \right)^{\alpha/2} \frac{\Gamma(n-\alpha/2)}{\Gamma(n)}, n>\alpha/2
  \end{equation}
\end{cor}
As expected, the $n^{\text{th}}$ dominant interference power expectation
decreases with path-loss parameter $\alpha$. 
$E[I_n]$ increases with the increasing reception angle $\phi$ and
node intensity $\lambda$. As a function of $n$, $E[I_n]$ decreases with
$\Gamma(n-\alpha/2)/\Gamma(n)$, whereas the expectation of the $n^{\text{th}}$
nearest node's distance increases with $\Gamma(n+1/2)/\Gamma(n)$ \cite{haenggi2005distances}.

\begin{cor}
  The CDF of the $n^{\text{th}}$ dominant
  interference power is obtained as
  \begin{equation}
    F_{I_n}^{\phi}(z) = \frac{\Gamma\left( n, \Lambda_2\left(z^{-1}\right) \right)}{\Gamma(n)},
  \end{equation}
  where $\Gamma(a,b)=\int_b^{\infty}x^{a-1} \exp\{-x\} dx$ is the upper
  incomplete gamma function.
\end{cor}

Let us define the partial accumulative interference power as
\begin{equation}
  \mathcal{I}(n)= \sum_{k=n+1}^{\infty}I_k=
  \sum_{k=n+1}^{\infty}z_k^{-1}, n= 1,2, \cdots.
\end{equation}

\begin{thm}\label{thm2}[Partial Accumulative Interference Power]
  In a PPP on $\mathbb{R}^2$ with intensity
  $\lambda$, the partial accumulative interference $\mathcal{I}(n)$
  excluding the first $n$ dominant
  interferers to a receiver with reception
  angle $\phi$ is characterized by its Laplace functional
  \begin{equation}\label{lapIzn}
    \mathcal{L}_{\mathcal{I}(n)}(s | z_n) = \exp\left\{ \frac{ \phi \lambda}{\alpha}
      \bar{h}_{{2}/{\alpha}} q_{z_n}(s) \right\},
  \end{equation}
  where ${q_{z_n}}(s)= s^{\frac{2}{\alpha}}\gamma\left( -\frac{2}{\alpha},\frac{s}{z_n} \right) +\frac{\alpha }{2}z_n^{2/\alpha} $.
\end{thm}

\begin{IEEEproof}
  \begin{align}
    &\mathcal{L}_{\mathcal{I}(n)}(s | z_n) = E\left[ \exp\left\{-s
      \sum_{k=n+1}^{\infty}z_k^{-1} \right\}
      \right] \nonumber \\
    & = E\left[ \prod_{k=n+1}^{\infty}\exp\left\{  -s
      z_k^{-1} \right\} \right] \nonumber \\
    & \stackrel{(a)}{=} \exp\left\{ \int_{z_n}^{\infty}\left( \exp\left\{-s z^{-1}\right\}-1 \right)
      \Lambda_2(dz) \right\} \nonumber \\
    \begin{split}
      &\stackrel{(b)}{=}\exp\left\{ \frac{ \phi \lambda}{\alpha}  \bar{h}_{{2}/{\alpha}} \left( \frac{\alpha z_n^{2/\alpha}}{2}+ s^{2/\alpha}\gamma\left(-\frac{2}{\alpha},\frac{s}{z_n}\right)
        \right) \right\},
    \end{split}
  \end{align}
  where $(a)$ uses the Laplace functional for PPP
  \cite{NET-006,haenggi2012stochastic} ($
  \mathcal{L}_{\Phi}(f) = \exp\{ \int_{0}^{\infty}\left( \exp\{-f \}-1 \right)
  \Lambda(dx)\}$, here $f$ is non-negative function on $\mathds{R}^2$) and $(b)$ substitutes
  Eq.~\eqref{intensity2final}.
\end{IEEEproof}

\begin{cor}[Average of Accumulate Interference]
  The expectation of $\mathcal{I}(n)$ (partial accumulate interference without first $n$
  dominant interference) in Poisson random network is
  \begin{equation}
    \bar{\mathcal{I}}(n)= \frac{2}{\alpha-2} \left(\frac{\bar{h}_{2/\alpha} \phi
        \lambda}{2} \right)^{\alpha/2} (n)_{1-\alpha/2}.
  \end{equation}
  where $(n)_{(\cdot)}$ is the Pochhammer sequence and $(n)_{1-\alpha/2}=\frac{\Gamma[n+1-\alpha/2]}{\Gamma[n]}$.
\end{cor}

\begin{IEEEproof}
  According to the definition of the Laplace functional, the first moment
  of $\bar{\mathcal{I}}(n)$ can be directly obtained by the derivative of its
  Laplace functional evaluated at $0$ on condition that $z_n$ is
  known:
  \begin{equation}\label{conditionalIzn1}
    E[\mathcal{I}(n)| z_n] = - \frac{\partial
      \mathcal{L}_{\mathcal{I}(n)}(s|z_n)}{\partial{s}}|_{s=0}.
  \end{equation}
  Since the derivative of $\mathcal{L}(s,z_n)$ can be directly
  calculated as
  \begin{align}\label{conditionalIzn2}
    &\frac{\partial \mathcal{L}_{\Ii(n)}(s|z_n)}{\partial{s}}\nonumber \\
    =&-
       \mathcal{L}_{\Ii(n)}(s|z_n)
       \int_{z_n}^{\infty}
       \frac{1}{z}
       e^{-s/z}
       \Lambda_2(dz)
       \nonumber \\
    =&-
       \mathcal{L}_{\Ii(n)}(s|z_n)\frac{
       \phi \lambda
       \bar{h}_{2/\al}}{\alpha}
       s^{2/\alpha-1}\gamma(1-2/\alpha,s/z_{n})
  \end{align}
  we have
  \begin{equation}
    \lim_{s \to 0} \frac{\partial \mathcal{L}_{\Ii(n)}(s|z_n)}{\partial{s}} = - \frac{\bar{h}_{2/\al}  \phi \lambda z_n^{2/\alpha-1}}{\alpha-2}.
  \end{equation}

  Thus we can calculated the average of accumulate interference as
  follows
  \begin{equation}\label{conditionalIzn3}
    \bar{\Ii}(n) = E\left[\Ii(n) \right] 
    = \int_0^{\infty} E\left[\Ii(n)| z_n\right]  f_{z_n}^{\phi}(z_n) dz_n.
  \end{equation}
  Applying Theorem~\ref{theorem1} and substituting
  Eq.~\eqref{conditionalIzn1} gives us the expectation of
  $\Ii(n)$
  \begin{equation}
    \bar{\Ii}(n) =\frac{2}{\alpha-2} \left(\frac{\bar{h}_{2/\al} \lambda
        \theta}{2}\right)^{\alpha/2} \frac{\Gamma(n+1-\alpha/2)}{\Gamma(n)}.
  \end{equation}

\end{IEEEproof}

\begin{cor}[Scaling from Omni-directional Reception Case]
  In a Poisson random network with density $\lambda$, the partial accumulative interference taken in
  angle $\phi$ directional reception averagely can be equivalent to the
  accumulative taken in omni-directional reception, if the
  interference within $B_0(R)$ is avoided, where
  \begin{equation}
    R = \left(\frac{\phi \bar{h}_{2/\al}}{2} \right)^{\frac{2}{\alpha(2-\alpha)}}
    \left( \frac{(n)_{1-\alpha/2}}{\pi \bar{h}}
    \right)^{\frac{1}{2-\alpha}} \lambda^{\frac{1}{\alpha}},
  \end{equation}
  here $\bar{h}=E[h]$. 
\end{cor}

\begin{IEEEproof}
  The accumulative interference without the nodes within $B_0(R)$ is
  formulated as
  \begin{equation}
    \Ii_{\Phi \backslash B_0(R)} = \sum_{x_i \in \Phi \backslash B_0(R)}
    h_i ||x_i||^{-\alpha}.
  \end{equation}
  The Laplace functional of $\Ii_{\Phi \backslash B_0(R)} $ is
  \begin{align}
    \mathcal{L}_{\Ii_{\Phi \backslash B_0(R)} }(s|R) &= E\left[\exp\{ -s
                                                       \Ii_{\Phi \backslash
                                                       B_0(R)}  \}\right]
                                                       \nonumber \\
                                                     & = E\left[\prod_{{\Phi
                                                       \backslash B_0(R)}}
                                                       E_h\left[
                                                       \exp\left\{ -s h_i ||x_i||^{-\alpha} \right\} \right]
                                                       \right] \nonumber
    \\
                                                     & = \exp\left\{ \int_{{\Phi \backslash B_0(R)}} \!\!\!\!\left( \!E_h\left[
                                                       e^{- s h ||x||^{-\alpha}} \right] -1 \right)\lambda dx  \right\}
  \end{align}
  Thus the derivative of $\mathcal{L}_{I_{\Phi \backslash B_0(R)}
  }(s|R)$ is
  \begin{align}
    &\frac{\partial}{\partial s}\mathcal{L}_{I_{\Phi \backslash B_0(R)}
      }(s|R) \nonumber \\
    &= \mathcal{L}_{I_{\Phi \backslash B_0(R)} }(s|R) \left\{ -\lambda
      \int_{{\Phi \backslash B_0(R)}}\!\!\!\!\!\!\!\!\!\! E_h\left[
      -h ||x||^{-\alpha}e^{- s h ||x||^{-\alpha}} \right] dx\right\}
  \end{align}
  Then we have
  \begin{equation}
    \!\!E\!\left[ \Ii_{\Phi \backslash B_0(R)} \right] \!=\!- \!\lim_{s \to 0} \frac{\partial}{\partial s}\mathcal{L}_{\Ii_{\Phi \backslash B_0(R)}}(s|R)=\! 2\pi \bar{h} \frac{R^{2-\alpha}}{\alpha-2}.
  \end{equation}        
  Then comparing $
  E\left[ \Ii_{\Phi \backslash B_0(R)} \right]$ with
  $\bar{\Ii}(n)$ gives the equivalent condition.
\end{IEEEproof}

\begin{cor}[Lower bound of accumulative interference]
  A lower bound on the Laplace functional of partial accumulative interference $\Ii(n)$
  is
  \begin{equation}
    \!\!\mathcal{L}^l_{\Ii(n)}\! (s)\!=\! exp\left\{\! n\!+\!\frac{\phi \lambda
        \bar{h}_{2/\al} s^{2/\alpha}}{\alpha} E_{z_n}\!\left[\gamma(-2/\alpha, s/z_n)\right] %
    \!\! \right\},
  \end{equation}
  where  $E_{z_n}\left[\gamma(-2/\alpha, s/z_n)\right]$ is the expectation of
  $\gamma(-2/\alpha, s/z_n)$ with the support of probability
  density function of $z_n$, i.e. $f_{z_n}^{\phi}$.
\end{cor}

\begin{IEEEproof}
  The lower bound is obtained straightforwardly by applying Jensen's
  inequality. Since exponential function $exp(x)$ is a convex function
  regarding $x$, thus we have
  \begin{align}
    &\int_{0}^{\infty} \mathcal{L}_{\Ii(n)}(s|z_n) f_{z_n}^{\phi}(z_n) d z_n
      \nonumber \\
    & \geq \exp \left \{\int_{0}^{\infty} \frac{ \phi \lambda \bar{h}_{2/\al}}{\alpha}
      q_{z_n} (s) f_{z_n}^{\phi}(z_n) d z_n  \right\} \nonumber \\
    & = \exp\left\{n + \frac{\phi \lambda
      \bar{h}_{2/\al} s^{2/\alpha}}{\alpha} E_{z_n}\left[\gamma(-2/\alpha, %
      s/z_n)\right] \right\}.
  \end{align}

\end{IEEEproof}

\begin{cor}[Upper bound of accumulative interference]
  An upper bound on the Laplace functional of partial accumulative interference $\Ii(n)$
  is
  \begin{equation}
    \mathcal{L}^u_{\Ii(n)} (s)= \exp \left\{ \frac{\phi \lambda \bar{h}_{2/\al}}{\alpha}
      \gamma(s,\bar{z}_n) \right\},
  \end{equation}
  where $\bar{z}_n = E[z_n] = \left( \frac{\bar{h}_{2/\al} \lambda \phi}{2}
  \right)^{-\alpha/2} \frac{\Gamma(n+\alpha/2)}{\Gamma(n)}.$
\end{cor}
\begin{IEEEproof}
  See Appendix~\ref{app_concavity}.
  
\end{IEEEproof}


%% file: section/sec-application.tex
\section{Applications to Communication Reliability}\label{sec:app}

In this section, communication reliability is studied with Nakagami-m
fading model, the channel gain $h$ has for PDF
\begin{equation}\label{h_pdf}
  f_h(x) = \frac{m^m x^{m-1}}{\Omega^m \Gamma(m)} \exp\left\{-\frac{m
  	x}{\Omega} \right\}, \hspace{12pt} m>\frac{1}{2},
\end{equation}
where $m$ is the fading parameter and $\Omega$
the fading power.

\subsection{Outage Probability}\label{outPro}
Outage of communication occurs when the Signal to
Interference Ratio (SIR) drops below a threshold $\eta$. The outage
probability can be calculated by evaluating the CDF of the SIR at the threshold $\eta$
\begin{equation}
  F_{\text{SIR}}^{\phi}(\eta)\hspace{-2pt} =\hspace{-2pt} \mathbb{P}\left(  \eta > \frac{h
      u^{-\alpha}}{\mathcal{I}(n)} \right) \hspace{-2pt}=\hspace{-3pt}E_{\mathcal{I}(n)}\left[
    F_h\left( u^{\alpha} \eta \mathcal{I}(n) \right) \right],
\end{equation}
where $u$ is the Euclidean distance between transmitter and receiver, $E_{\mathcal{I}(n)}[\cdot]$ denotes expectation regarding
$\mathcal{I}(n)$ and
\begin{equation}
  F_h(x)\stackrel{(a)}{=} \frac{\gamma(m,m x/ \Omega)}{\Gamma(m)} \stackrel{(b)}{=} 1-
  \sum_{k=0}^{m-1} \frac{\left( m x/\Omega \right)^k}{k!}\exp\{-m x/\Omega \} ,
\end{equation}
where $(a)$ uses PDF \eqref{h_pdf} and $(b)$ achieves when $m$ is
positive integer. Then applying Eq.~\eqref{pdfzn} and
\eqref{lapIzn} gives
\begin{align}\label{sirCDF}
  &\hspace{-3pt}F_{\text{SIR}}^{\phi}(\eta) = E_{\mathcal{I}(n)}\left[1- \sum_{k=0}^{m-1}
                         \frac{\left( m u^{\alpha} \eta \mathcal{I}(n) \right)^k}{\Omega^k k!} e^{-\frac{m
                         	u^{\alpha} \eta \mathcal{I}(n)
                         }{\Omega}} \right] \nonumber \\
                       & =1-\! \sum_{k=0}^{m-1}
                         \frac{\left( m u^{\alpha} \eta \right)^k}{\Omega^k k!}
                         E_{\mathcal{I}(n)}\left[  \mathcal{I}^k(n) e^{-\frac{m u^{\alpha}
                         \eta \mathcal{I}(n)}{\Omega}}\right] \nonumber \\
                       & = 1-\sum_{k=0}^{m-1}
                         \frac{\left( -m u^{\alpha} \eta \right)^k}{\Omega^k k!}
                         E_{z_n}\left[ \mathcal{L}_{\mathcal{I}(n)}^{(k)}\left(\left.\frac{m u^{\alpha} \eta}{\Omega}  \right\rvert z_n\right) \right],
\end{align}
where $\mathcal{L}_{\mathcal{I}(n)}^{(k)}(\cdot|z_n)$ is the $k_{th}$
derivative of $\mathcal{L}_{\mathcal{I}(n)}(\cdot|z_n)$.

In the Nakagami-m fading case, 
\begin{equation}
  \bar{h}_{2/\alpha}=E_h[h^{2/\alpha}]= \left( \frac{m}{\Omega}
  \right)^{-\frac{2}{\alpha}} \frac{\Gamma(m+\frac{2}{\alpha})}{\Gamma(m)}.
\end{equation}

\subsection{Error probability under QoS Constraint}
QoS (Quality of Service) is defined by parameter pair $(\epsilon_q, Q_{\max})$ and used to measure communication link quality, where $Q_{\max}$ is the
tolerable queue length for service data at transmitter and $\epsilon_q$ is the
violation probability of constraint $Q_{\max}$.
\cite{wu2003effective,chang1995effective} give the approximation of
$\epsilon_q$ as
\begin{equation}
  \epsilon_q \sim \exp\left\{- \theta Q_{\max}\right\} ,
\end{equation}
where $\theta$ is QoS exponent\footnote{Larger $\theta$ stands for higher
QoS requirement, i.e. smaller $Q_{\max}$ or violation probability bound $\epsilon_q$.}. 
For any required QoS, corresponding effective bandwidth $a(\theta)$
\cite{wu2003effective,chang1995effective} gives the minimal data rate to meet
the QoS requirement $(Q_{\max},\epsilon_q)$, defined as
\begin{equation}
  a(\theta) = \lim_{t \to \infty} \frac{\log E[\exp\left\{\theta
  	A(t) \right\}  ]}{t \theta},
\end{equation}
where $A(t)$ is the cumulative source data over time interval
$[0,t)$. If transmitter can send data out with guaranteed rate 
$r=\alpha(\theta)$, violation error probability can be bounded by $\epsilon_q$.
However,
data rate over wireless channel is dynamic and unreliable. The
selected rate $r$ by transmitter could be failed due to poor
SIR. With derived result in Section~\ref{outPro}, the error
probability that wireless channel can not provide rate $r$ is 
\begin{equation}
  \epsilon_r=\mathbb{P}[\log(1+\text{SIR})<a(\theta)] =
  F_{\text{SIR}}^{\phi}(\exp\{a(\theta)\}-1).
\end{equation}
The overall error probability $\epsilon$ is due to
either queue violation either  channel fading and can be approximated as
\begin{align}\label{total_err}
  \epsilon \approx 1-(1-\epsilon_q)(1-\epsilon_r)
  =\epsilon_q + \epsilon_r -\epsilon_q \epsilon_r.
\end{align}
Hence for a given queue length constraint $Q_{\max}$ and
chosen transmission $r$, the total error can be
approximated by Eq.~\eqref{total_err}.

\subsection{Relationship between $r$ and $\epsilon$}
The following theorem shows the whether a given QoS specification is possible:
\begin{thm}\label{theoremQos}
  Assume a wireless link with error probability $\epsilon_r(r)$ for
  corresponding link achievable rate $r$. Denote the target QoS specification by
  ($\epsilon^{\prime}, Q_{max}$). The target QoS is possible to be met by rate
  adaptation (increasing $r$), if the condition
  \begin{equation}\label{theoremQos1}
    \epsilon_r(r^{\ast}) \leq 1- \sqrt[2]{1-\epsilon^{\prime}}
  \end{equation}
  is met. Here $r^{\ast}$ is the root to equation
  \begin{equation}\label{theoremQos2}
    \epsilon_q(r) = \epsilon_r(r),
  \end{equation}
  where $\epsilon_q(r)$ is the queue violation probability with
  service rate $r$ and maximum tolerable queue length at transmitter
  $Q_{max}$. Note that the equation $\epsilon_q(r) =
  \epsilon_r(r)$ has at most one root.
\end{thm}

\begin{IEEEproof}
  As stated the total error $\epsilon(r)$ is actually a function of the
  selection rate $r$, which can be formed as
  \begin{equation}
    \epsilon(r) = \epsilon_q(r) + \epsilon_r(r)
    -\epsilon_q(r)\epsilon_r(r).
  \end{equation}
  For an error probability $\epsilon^{\prime}$, the selected rate
  must satisfy
  \begin{equation}
    r> a(\theta^{\prime}),
  \end{equation}
  where
  \begin{equation}
    \theta^{\prime}=-\log{\epsilon^{\prime}}/Q_{max},
  \end{equation}
  otherwise the packet queue at transmitter would not be stable and there
  would be no bound on the queue violation error.

  We rewrite $\epsilon(r)$ and have
  \begin{align}\label{minErrIneq}
    \epsilon(r) &= 1- \left( 1- \epsilon_r(r) \right) \left(
                  1-\epsilon_q(r) \right) \nonumber \\
                &\stackrel{(a)}{\geq}  1- \left( 1- \frac{\epsilon_r(r) +
                  \epsilon_q(r)}{2} \right)^{2},
  \end{align}
  where $(a)$ uses that arithmetic mean of non-negative numbers is
  greater than or equal to their geometric mean. The equality
  achieves when the $\epsilon_r(r) = \epsilon_q(r)$.
  By setting $a(\theta) = r$, we have
  \begin{equation}\label{errq1d}
    \frac{ \partial{\epsilon_q(r)}}{\partial{r}} = - Q_{max} e^{-
      \theta Q_{max}} \frac{\partial{\theta}}{\partial{r}}.
  \end{equation}
  According to \cite{chang1995effective},
  effective bandwidth is an increasing function of $\theta$, thus we
  have
  \begin{equation}\label{errq2d}
    \frac{\partial{r}}{\partial{\theta}}> 0.
  \end{equation}

  Combining Eq.~\eqref{errq1d} and \eqref{errq2d} gives
  \begin{equation}
    \frac{ \partial{\epsilon_q(r)}}{\partial{r}}< 0.
  \end{equation}
  Thus, we concludes that $\epsilon_q(r)$ is monotonically decreasing
  function of selected rate $r$, i.e. selecting larger rate $r$ ($r
  \geq a(\theta^{\prime})$) leads smaller queue violation error
  $\epsilon_q(r)$.

  On the other hand, we would like to show that communication link
  failure error $\epsilon_r(r)$ is an increasing function of selected
  rate $r$. Assuming that the probability density function of SINR is
  $f$ ($f>0$ in its domain) and using Shannon capacity, we have
  \begin{align}
    \frac{\partial{\epsilon_r(r)}}{\partial{r}} &=
                                                  \frac{\partial{}}{\partial{r}}
                                                  \int_{0}^{e^r -1}
                                                  f(x) dx \nonumber \\
                                                &= f(e^r-1)e^r
                                                  \nonumber \\
                                                & > 0.
  \end{align}

  Since $\epsilon_r(r)$ is monotonically increasing and $\epsilon_q$
  is monotonically decreasing when we choosing larger $r$, there is
  one and only one root to $\epsilon_r(r) = \epsilon_q(r)$ if the
  condition
  \begin{equation}\label{edgecondition}
    \epsilon_r\left( a(\theta^{\prime}) \right) \leq \epsilon_q\left(
      a(\theta^{\prime}) \right)
  \end{equation}
  is met.

  Assume that the condition~\eqref{edgecondition} is met and
  $r^{\prime}$ is the unique root to $\epsilon_r(r) = \epsilon_q(r)$,
  according to \eqref{minErrIneq}, we get
  
  \begin{align}\label{errInf}
    \inf{\epsilon} &= 1- \left( 1-
                     \frac{\epsilon_r(r^{\ast}) + \epsilon_q(r^{\ast})}{2}
                     \right)^{2} \nonumber \\
                   &= 1- \left( 1- \epsilon_r(r^{\ast}) \right).
  \end{align}

  We have the QoS requirement that error is no larger than
  $\epsilon^{\prime}$. Thus, this is possibly be met by choosing
  better communication rate if 
  \begin{equation}\label{infCom}
    \inf{\epsilon} \leq \epsilon^{\prime}.
  \end{equation}
  Otherwise, we can not guarantee that the initial QoS specification
  $(\epsilon^{\prime}, Q_{max})$ would be met by choosing larger $r$.

  Substituting Eq.~\eqref{errInf} into \eqref{infCom} gives the
  condition
  \begin{equation}
    \epsilon_r(r^{\ast}) \leq 1- \sqrt[2]{1-\epsilon^{\prime}},
  \end{equation}
  where $\epsilon_r(r^{\ast}) = \epsilon_q(r^{\ast})$.
\end{IEEEproof}

Theorem~\ref{theoremQos} gives the sufficient condition to evaluate if
a proposed QoS specification can be reasonably fulfilled with a
certain communication link condition. However, it is also possible
that we can find a root $r^{\ast}$ for Eq.~\eqref{theoremQos2}
that can not fulfill the condition of inequatlity
\eqref{theoremQos1}. In this case, we claim that it is possible to find
a rate $r^{\ast}$ that brings lowest total error $\epsilon$ but
without meeting the QoS requirement $(\epsilon^{\prime},
Q_{max})$. Still, this rate selection $r^{\ast}$ gives the smallest
error, i.e. $\epsilon(r^{\ast})$.

There is worse situation where $\epsilon_r(a(\theta^{\prime})) >
\epsilon_q^{\prime}$. In this case, there is no way for QoS $(\epsilon^{\prime},
Q_{max})$ to be met. Due to the monotonic property of $\epsilon_r$ and
$\epsilon_q$ regarding $r$, Eq.~\eqref{theoremQos2} does not has
root. But it is still possible find rate $r$ such that $\epsilon(r) <
\epsilon(a(\theta^{\prime}))$. This could be done by solving the first
derivative equation
\begin{equation}
  \frac{\partial{\epsilon(r)}}{\partial{r}} = 0.
\end{equation}
subject to
\begin{equation}
  r>a(\theta^{\prime}).
\end{equation}


%% file: section/sec-numerical.tex
\section{Numerical Results}\label{sec:numResult}

This section shows some numerical results (``Sim.'') and their
comparisons with analytic results (``Ana.'').
We set the parameters as node intensity $\lambda=10^{-4}$, path-loss
exponent $\alpha=3$, $m=2$, $\phi=\pi/4$ and $\Omega=1$, unless stated otherwise.

\begin{figure}
  \centering
  \includegraphics[scale = 0.305]{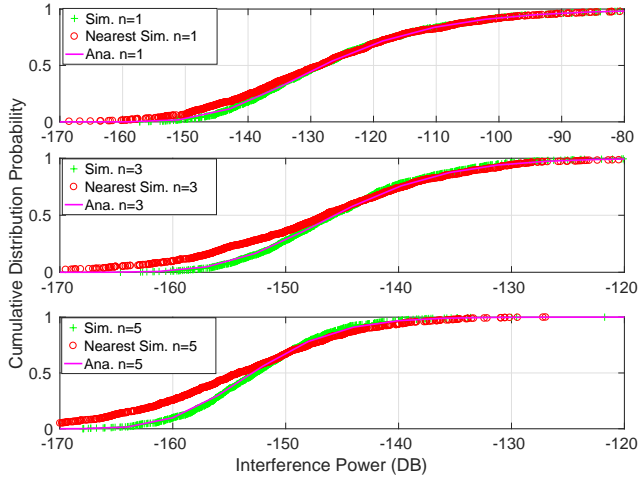}\vspace{-0.5cm}
  \caption{Cumulative Probability Function
    $F_{I_n}^{\phi}$ compared with $n^{\text{th}}$ nearest node's
    interference power CDF distribution. 
  }\label{fig:cpdf}\vspace{-0.5cm}
\end{figure}

As shown in Fig.~\ref{fig:cpdf}, curves of CDF for different $n^{\text{th}}$ dominant
interference power $I_n$ are given numerically and analytically. 
Since $I_n$ is directly sorted by the
interference power, we compare it with the $n^{\text{th}}$ nearest\footnote{Here
  $n^{\text{th}}$ nearest node is sorted out by Euclidean distance. Thus the $n^{\text{th}}$ nearest
interferer is the $n^{\text{th}}$ closest neighbor by distance.} interferer's
power (``Nearest Sim.'') under same fading context. As shown, the
difference between $F_{I_n}^{\phi}$ and distribution of ``Nearest Sim. n'' is larger
as $n$ increases. In small range of $I_n$, $F_{I_n}^{\phi}$ is smaller
than CDF sorted by distance. But in large range of $I_n$,
$F_{I_n}^{\phi}$ is larger. That means that distance-based
approximation can be overestimation or underestimation depending on
$I_n$. For
larger $n$, the Euclidean-distance-based approximation has larger
bias.

\begin{figure}
  \centering
  \includegraphics[scale = 0.3]{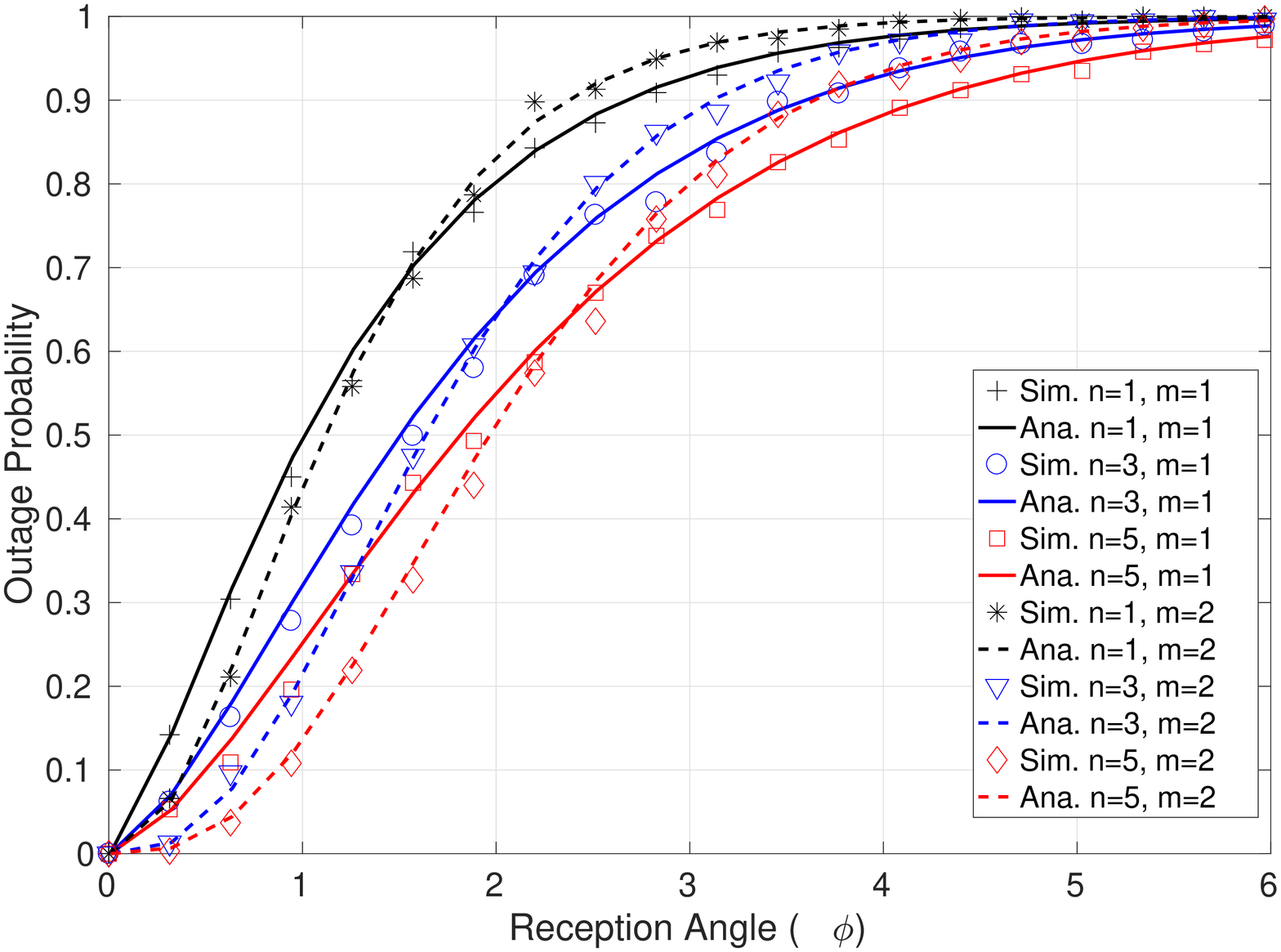}\vspace{-0.4cm}
  \caption{Outage probability ($F_{\text{SIR}}^{\phi}$) against reception
    angle $\phi$. }\label{fig:partialAngle}\vspace{-0.6cm}
\end{figure}

Fig.~\ref{fig:partialAngle} shows
the outage probability against the reception angle $\phi$, which
matches well with $F_{\text{SIR}}^{\phi}$ in Eq.~\eqref{sirCDF}. Here $\eta =1$. As expected, reception with larger angle $\phi$ is subject to heavier interference and thus the outage
probability increases along with the rising $\phi$. Additionally,
$F_{\text{SIR}}^{\phi}$ decreases obviously for increasing $n$, when excluding more dominant
interferers. The changes of outage probability vary with
network setting such as $\phi$. The Rayleigh fading ($m=1$) is
simulated for comparison. 
In small range of $\phi$, the outage probability of Rayleigh fading is
larger than that of $m=2$, since signal fading of interest due to
absence of direct line of sight (LOS) dominates. But, in large value of $\phi$, outage of Nakagami-m
($m=2$) is larger since LOS advantage makes receiver suffer more interference
as large $\phi$ leads to large number of interferers with LOS.

\begin{figure}
  \centering
  \includegraphics[scale = 0.3]{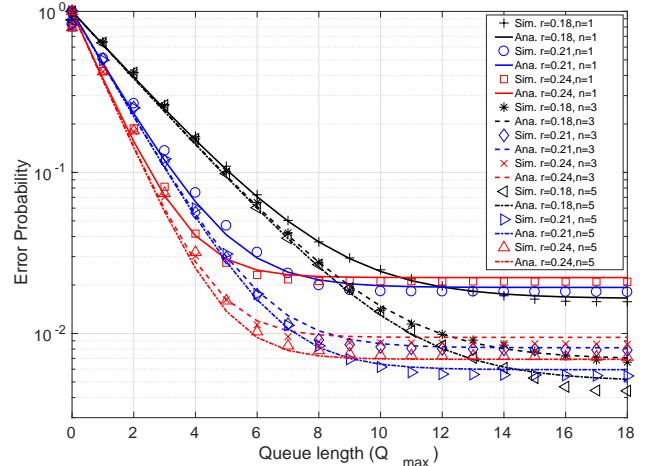}\vspace{-0.4cm}
  \caption{Data error ($\epsilon$) due to fading and queue violation. }\label{fig:dateLost}\vspace{-0.6cm}
\end{figure}

Fig.~\ref{fig:dateLost} shows total communication error that
matches well with Eq.~\eqref{total_err}. Poisson arrival with
parameter $\lambda_d=0.14$ is used and thus effective bandwidth is
$\alpha(\theta)= \lambda_d (\exp\{\theta\}-1)/{\theta}$. For deterministic rate
$r$ choice, total error $\epsilon$ decreases with increase of
$Q_{\max}$ in small range of $Q_{\max}$ but flattening
eventually in large range of $Q_{\max}$. This means smaller error can
be expected by loosening constraint on $Q_{\max}$ when initial $Q_{\max}$ is not large. Otherwise $\epsilon$
flattens around $\epsilon_r$. Choosing
larger $r$ is an effective way to get lower error before $\epsilon$
flattens but it could bring larger error in larger range of
$Q_{\max}$. However, larger $n$ could lead smaller total error even in
larger range of $Q_{\max}$.

%% file: section/sec-conclusion.tex
\section{conclusion}\label{sec:conclusion}

In this paper, we studied the dominant interference power in random
networks modeled by PPP. Both the $n^{\text{th}}$ dominant and partial
accumulative interference power were studied. We showed the bias of
Euclidean-distance-based approximation by the $n^{\text{th}}$ nearest interferer numerically. This bias could
be large for large $n$. Then, the obtained results
were used to evaluate communication link reliability by metrics of outage
probability and error probability with consideration of queue length
violation. The possible way to decrease outage probability and total
error was simulated and discussed.


%% file: section/sec-appendix.tex
\section{Upper bound of $\mathcal{L}_{\Ii(n)}(s|z_n)$}\label{app_concavity}
\begin{IEEEproof}
  For the upper bound of $\mathcal{L}_{\Ii(n)}(s)$, we will show that $\mathcal{L}_{\Ii(n)}(s|z_n)$
  is concave regarding $z_n$ asymptotically for growing $z_n$
  (decreasing $n_{th}$ dominant power), \emph{i.e.} the
  concavity of $\mathcal{L}_{\Ii(n)}(s|z_n)$ as a function of $z_n$ is also
  preserved for non-dense networks (small $\lambda$). If this
  condition satisfied, the conditional upper bound is as follows
  \begin{equation}
    \int_{0}^{\infty} \mathcal{L}_{\Ii(n)}(s|z_n) f_{z_n}^{\phi}(z_n) d z_n
    \leq \exp \left\{ \frac{\phi \lambda \bar{h}_{2/\al}}{\alpha}
      \gamma(s,\bar{z}_n) \right\},
  \end{equation}
  where
  \begin{equation}
    \bar{z}_n = E[z_n] = \left( \frac{\bar{h}_{2/\al} \lambda \phi}{2}
    \right)^{-\alpha/2} \frac{\Gamma(n+\alpha/2)}{\Gamma(n)}.
\end{equation}
  According to Theorem~\ref{thm2}, the Laplace
  functional fo partial accumulative interference is
  \begin{equation}
    \Ll_{\Ii(n)}(s|z_n) = exp\left\{ \frac{ \phi \lambda \bar{h}_{2/\al}}{\alpha}
      \gamma(s,z_n) \right\}.
  \end{equation}

  The second derivative of $\Ll_{\Ii(n)}(s|z_n)$ regarding to $z_n$ is
  \begin{align}\label{2edD}
    &\frac{\partial^{2}{\Ll_{\Ii(n)}(s|z_n)}}{\partial{z_n^{2}}}
      \nonumber \\
    =& \Ll_{\Ii(n)}(s| z_n)
       \left(  \frac{ \phi
       \lambda
       \bar{h}_{2/\al}}{\alpha}\right)^2
       \left(
       \frac{\partial{q_{z_n}(s)}}{\partial{z_n}}
       \right)^2 \nonumber \\
    & + \Ll_{\Ii(n)}(s| z_n)
      \frac{ \phi
      \lambda
      \bar{h}_{2/\al}}{\alpha} \frac{\partial^2{q_{z_n}(s)}}{\partial{z_n^2}}
      \nonumber \\
    = &\Ll_{\Ii(n)}(s| z_n)\frac{ \phi\lambda\bar{h}_{2/\al}}{\alpha} \left( \frac{ \phi
        \lambda
        \bar{h}_{2/\al}}{\alpha} \left(
        \frac{\partial{q_{z_n}(s)}}{\partial{z_n}}
        \right)^2 \!\!+\!   \frac{\partial^2{q_{z_n}(s)}}{\partial{z_n^2}}\right)
  \end{align}

  It is obvious that $\Ll_{\Ii(n)}(s| z_n)\frac{
    \phi\lambda\bar{h}_{2/\al}}{\alpha}$ is positive. Then it is the formula inside the
  parenthese that decides the sign of second derivative of
  $\Ll_{\Ii(n)}(s| z_n)$ in Eq.~\eqref{2edD}. Thus we need to
  analyze the two terms inside the parenthesis to see its sign. The $\frac{ \phi \lambda
    \bar{h}_{2/\al}}{\alpha} \left( \frac{\partial{q_{z_n}(s)}}{\partial{z_n}}
  \right)^2$ is positive. Since
  \begin{equation}
    q_{z_n}(s) = s^{2/\alpha} \gamma\left(-2/\alpha,\frac{s}{z_n}\right)
    +\frac{\alpha z_n^{2/\alpha}}{2},
  \end{equation}
  we have the first derivative of $q_{z_n}(s)$ regarding $z_n$ as
  \begin{equation}
    \frac{\partial{q_{z_n}(s)}}{\partial{z_n}} = \left( 1- e^{-s/z_n}
    \right) z_n^{-1+2/\alpha}
  \end{equation}
  and the second derivative as
  \begin{equation}
    \frac{\partial^2{q_{z_n}(s)}}{\partial{z_n^2}}\! =\! z_n^{-2-\alpha+\frac{2}{\alpha}}
    e^{-\frac{s}{z_n}}\! \left[ -\!\left( e^{\frac{s}{z_n}}\!-\!1 \right) z (\alpha \!-\!2) - s
      \alpha \right].
  \end{equation}

  Since $\alpha > 2$, and both $s$ and $z_n$ are positive, it is
  straightforward that $\frac{\partial^2{q_{z_n}(s)}}{\partial{z_n^2}} $ is
  always negative. On the other hand, as we mentioned that the $\left(
    \frac{\partial{q_{z_n}(s)}}{\partial{z_n}} \right)^2$ is positive, we
  could hardly decide if Eq.~\eqref{2edD} is positive or negative
  directly. Calculating the the derivative of Eq.~\eqref{2edD}
  would make the analysis more complicated since the root of it is not
  be analytically obtained. Thus we use the power series for its
  asymptotic analysis: 
  \begin{equation}
    exp(z) = \sum_{k=0}^{\infty} \frac{z^k}{k!}.
  \end{equation}
  
  Applying this, we have
  \begin{align}\label{asy1}
    &\left(\frac{\partial{q_{z_n}(s)}}{\partial{z_n}} \right)^2
      \nonumber \\
    =& \left( 1-e^{-s/z_n}\right)^2 z_n^{-2+4/\alpha} \nonumber \\
    =& \left( 1-2 e^{-s/z_n} + e^{-2s/z_n}\right) z_n^{-2+4/\alpha}
       \nonumber \\
    =& \left( 1-2 \sum_{k = 0}^{\infty} \frac{(-s)^k}{k!} z_n^{-k} +
       \sum_{k = 0}^{\infty} \frac{(-ss)^k}{k!} z_n^{-k} \right)
       z_n^{-2+4/\alpha} \nonumber \\
    =& \left( \sum_{k = 1}^{\infty} \frac{\left[ (-2s)^k - 2 (-s)^{k}\right]}{k!} z_n^{-k} \right)
       z_n^{-2+4/\alpha} \nonumber \\
    =& z_n^{2/\alpha-2} \sum_{k = 2}^{\infty} \frac{\left[ (-2s)^k - 2
       (-s)^{k}\right]}{k!} z_n^{2/\alpha-k} \nonumber \\
    =& z_n^{2/\alpha -2} \mathcal{O}(z^{2/\alpha-2}).
  \end{align}

  Similarly, the second derivative of $q_{z_n}(s)$ can be formed as
  \begin{align}\label{asy2}
    &\frac{\partial^2{q_{z_n}(s)}}{\partial{z_n^2}} \nonumber \\
    =& z_n^{\frac{2}{\alpha} -2 } \left(- \left( 1-\frac{2}{\alpha} \right) +
       e^{-s/z_n} \left( 1- \frac{2}{\alpha} -\frac{s}{z_n} \right)\right) \nonumber \\
    =& z_n^{\frac{2}{\alpha} -2 } \!\left(\!\left(\! 1\!-\!\frac{2}{\alpha} \right)\! \left(  \sum_{k=0}^{\infty} \frac{(-s)^k}{k!}
       z_n^{-k} \!-\!1\right) \!- \!\frac{s}{z_n} \sum_{k=0}^{\infty} \frac{(-s)^k}{k!}
       z_n^{-k} \right) \nonumber \\
    =& z_n^{\frac{2}{\alpha} -2 }\left( \left( 1-\frac{2}{\alpha} \right)\sum_{k=1}^{\infty} \frac{(-s)^k}{k!}
       z_n^{-k} -s \sum_{k=0}^{\infty} \frac{(-s)^k}{k!} z_n^{-k-1}
       \right) \nonumber \\
    =& z_n^{2/\alpha -2} \mathcal{O}(z_n^{-1}).
  \end{align}

  Substituting Eq.~\eqref{asy1} and \eqref{asy2} into \eqref{2edD}
  gives 
  \begin{align}\label{2edDasym}
    &\frac{\partial^{2}{\Ll_{\Ii(n)}(s|z_n)}}{\partial{z_n^{2}}}
      \nonumber \\
    =& \Ll_{\Ii(n)}(s| z_n)
       \frac{ \phi \lambda \bar{h}_{2/\al}}{\alpha z_n^{2-2/\alpha}} \left(\! \frac{ \phi
       \lambda \bar{h}_{2/\al} }{\alpha}
       \!\mathcal{O}(z_n^{2/\alpha -2}) \!+ \!\mathcal{O}(z_n^{-1}) \!\right).
  \end{align}

  Since $\alpha >2$, $2-\frac{2}{\alpha} > 1$. Then
  $\mathcal{O}(z_n^{-1})$ fades slower than $\mathcal{O}(z_n^{2/\alpha
    -2})$ as $z_n$ increases. Also
  \begin{align}
    \lim_{z_n \to 0} \frac{ \phi \lambda \bar{h}_{2/\al} }{\alpha} \left(
    \frac{\partial{q_{z_n}(s)}}{\partial{z_n}}\right)^2  &= 0,
                                                           \nonumber \\
    \lim_{z_n \to 0}\frac{\partial^2{q_{z_n}(s)}}{\partial{z_n^2}}  &= 0.
  \end{align}

  Thus, the positive term in Eq.~\eqref{2edDasym} approaches to
  $0$ faster than the negative term. We can say when $z_n$
  is larger than a certain value $z$,
  $\frac{\partial^{2}{\Ll_{\Ii(n)}(s|z_n)}}{\partial{z_n^{2}}}$ remains
  negative. In addition, for non-dense networks, i.e. $\lambda$ is
  small, positive term of
  $\frac{\partial^{2}{\Ll_{\Ii(n)}(s|z_n)}}{\partial{z_n^{2}}}$ is
  relatively small and then second derivative of $\Ll_{\Ii(n)}(s|z_n)$ is
  negative throughout positive axis of $\mathbb{R}$.
\end{IEEEproof}
